 \documentstyle[prl,aps]{revtex}  
 \def\ket{\rangle} 
 
\begin{document} 
\draft 
 
 
 \twocolumn[\hsize\textwidth\columnwidth\hsize  
 \csname @twocolumnfalse\endcsname              
 
\title{Efficient Scheme for Initializing a Quantum Register \\  
with an Arbitrary Superposed State} 
\author{Gui-Lu Long$^{1,2,3}$ and Yang Sun$^{1,2,4}$} 
\address{  
$^1$Department of Physics, Tsinghua University, Beijing 100084, 
P.R. China\\ 
$^2$Laboratory for Quantum Information and Measurement, Key Lab of MOE,  
Tsinghua University, Beijing 100084, P.R. China\\ 
$^3$Institute of Theoretical Physics, Chinese Academy of Sciences, 
Beijing 100080, P.R. China\\ 
$^4$Department of Physics and Astronomy, University of Tennessee, Knoxville,  
TN 37996} 
 
\date{\today} 
\maketitle 
 
\begin{abstract} 
Preparation of a quantum register is an important step  
in quantum computation and quantum information processing.  
It is straightforward to build a simple quantum state such as $|i_1 i_2  
\cdots i_n\ket$ with $i_j$ being either 0 or 1, 
but is a non-trivial task  
to construct an {\it arbitrary} superposed quantum state.  
In this Letter, we present a  
scheme that can most generally  
initialize a quantum register with an arbitrary superposition 
of basis  
states. Implementation of this scheme requires $O(Nn^2)$  
standard 1- and 2-bit gate operations, {\it without introducing additional 
quantum bits}. Application of the 
scheme in some special cases is discussed. 
\end{abstract} 
 
 
 ]  

Research on quantum computers and quantum information processing  
has been a fast developing interdisciplinary field over the  
past years. As a new branch of science overlapping quantum physics and 
classical information theory, it 
resembles in some ways both of the subfields, but differs from  
each of them in 
many other respects.   
In quantum computation and quantum information processing,  
the concept of quantum superposition of basis states $|i_1 i_2 
\cdots i_n\ket$  
is used and massive parallelism is achieved \cite{r1}. For instance, 
a significant speed-up  
over classical computers, at least theoretically,  
has been gained in prime-factorization \cite{r2} and 
quantum searching \cite{r3}.  
Nevertheless,
some simple operations for a classical computer can not be easily  
implemented in a quantum  
computer. A vivid example is the need of introducing the  
quantum error correction scheme   
to overcome the decoherence problem in quantum computers.  
This has been obtained with admirable genius \cite{r4} 
whereas the corresponding classical coding scheme is straightforward. 
 
Quantum computing is realized by quantum gate operations. It 
has been  
shown that a finite set of basic gate operations can be used to construct
any 
quantum computation gate operation \cite{deu89}. 
This universality of quantum computation 
has been studied by many authors \cite{div,bar,lloyd}.
A quantum 
circuit, which is a network of  
gate operations for certain purpose, has been constructed, 
for example, for basic arithmetic  
\cite{rmath} and efficient factorization \cite{rshor}.  
 
Initializing a quantum register to an {\it arbitrary} superposition of  
basis  
states is a seemingly simple, yet difficult problem.  
Addition of two numbers in a classical computer could not be 
easier, but  
addition of two quantum states $a_1|\psi_1\ket + a_2|\psi_2\ket$ 
is not easy  
at all. However, superposition is the basic ingredient in quantum 
computing and  
quantum information processing.  
An efficient scheme for initializing an arbitrary   
superposition  
for a quantum register is very much desired. 
An efficient scheme for initializing a quantum  
register for a known function of amplitude distribution was given 
by Ventura and Martinez (VM)   
with $n+1$ additional quantum bits (qubits) \cite{ventura}.
 
In this Letter, we present a general scheme that  
initializes a quantum register without introducing additional 
qubits. For some  
quantum computing tasks, introduction of additional qubits is not permitted.  
Thus our scheme may be appreciated by these circumstances.  
Furthermore, qubits are a precious resource in practice, and any saving 
is a great relief for existing technology, especially at the present time when 
researchers are  
striving to make more qubits available. 
 
Starting with the state $|0\cdots 0\ket$,  
we want to transform  
this state  
to a general superposed state having the form 
\begin{eqnarray} 
|\psi\ket=\sum_{i=0}^{N-1} a_i |i\ket. 
\label{e1} 
\end{eqnarray} 
Normalization of this state vector is assumed. 
The coefficients $a_i$ are in general complex numbers with the requirement 
$|a_i|\le 1$.  
Here, $i$ is a short notation for a set of indices $\{i_1 i_2   
\cdots i_j \cdots i_n\}$ 
with $n=\log_2N$ being the total number of qubits in the register,  
and $i_j$ denotes the two possible states (0 or 1) of the $j$th qubit.  
To be concrete, our notation implies  
\begin{displaymath} 
i = \left\{  
\begin{array}{rcl} 
0 & \longrightarrow & \{00\cdots 00\} \\ 
1 & \longrightarrow & \{00\cdots 01\} \\ 
2 & \longrightarrow & \{00\cdots 10\} \\ 
\vdots &&\\ 
N-1 & \longrightarrow & \{11\cdots 11\}  
\end{array} \right .  
\end{displaymath} 
Thus, $|\psi\ket$ in Eq. (\ref{e1}) is a general quantum  
superposition of $N$ basis states, 
and each of the basis states is a product state of $n$ qubits.    
 
Our scheme involves only two kinds of elementary unitary transformations,  
or gate operations. The first kind of 
gate operation is a single-bit 
rotation $U_{\theta}$, 
\begin{eqnarray} 
U_{\theta}\left[ 
\begin{array} {c} 0 \\ 1 \end{array} 
\right]= 
\left[ 
\begin{array}{rr} \cos\theta & \sin\theta \\ \sin\theta & -\cos\theta 
\end{array} 
\right]\left[ 
\begin{array}{c} 0 \\ 1\end{array} 
\right]. 
\label{rotation} 
\end{eqnarray} 
It differs slightly from an ordinary rotation because  
it is an ordinary rotation for the $|0\ket$ part only, but has a minus sign  
for the $|1\ket$ part.  
Upon operation, a qubit in the state $|0\ket$  
is transformed into a superposition  
in the two state: $(\cos\theta, \sin\theta)$. Similarly, a qubit in the state  
$|1\ket$ is transformed into $(\sin\theta, -\cos\theta)$.  
It is useful to identify some special cases in Eq. (\ref{rotation}).  
When $\theta=0$, it does not change 
$|0\ket$, but converts the sign of the state $|1\ket$.  
When $\theta={\pi \over 4}$, 
$U_{\theta}$ is reduced to   
the Hadamard-Walsh transformation \cite{hw}.  
Finally, when $\theta={\pi \over 2}$, it 
serves as the NOT operation:  
it changes  
$|0\ket$ to $|1\ket$, and $|1\ket$ to $|0\ket$. 
 
The second kind of gate operation is  
the controlled$^k$-operations.  
As illustrated below, it is an operation that has a string of  
$k$ controlling qubits:  
 
\begin{picture}(50,40) 
\put(10,14){\framebox(12,12){$i_1$}} 
\put(22,20){\line(1,0){18}} 
\put(40,14){\framebox(12,12){$i_2$}} 
\put(52,20){\line(1,0){12}} 
\put(76,20){\makebox(1,1){$\cdots$}}  
\put(88,20){\line(1,0){12}} 
\put(100,14){\framebox(12,12){$i_k$}} 
\put(112,20){\line(1,0){22}} 
\put(140,20){\circle{12}} 
\put(140,20){\makebox(1,1){\small $U_\alpha$}}  
\end{picture} 
 
\noindent 
The squares represent the   
controlling qubits, and the circle is a unitary operation on the target 
qubit. The  
operation is a conditional one that  
is activated only when the controlling qubits hold the respective 
values indicated in the squares. Controlled$^k$-operations
can be constructed by 
$O(k^2)$ standard 1- and 2-bit gate operations \cite{bar}.
 
With these basic gate operations at our disposal, 
we now proceed from simple examples to the most general case.  
For a 2-qubit system, the transformation can be expressed as 
\begin{eqnarray} 
|00\ket  
&\rightarrow&\sqrt{|a_{00}|^2 + |a_{01}|^2} |00\ket 
                +\sqrt{|a_{10}|^2 + |a_{11}|^2} |10\ket \nonumber\\ 
&\rightarrow&|0\ket \left[ a_{00}|0\ket + a_{01}|1\ket \right]   
                +|1\ket \left[ a_{10}|0\ket + a_{11}|1\ket \right] \nonumber\\  
&=& a_{00}|00\ket + a_{01}|01\ket + a_{10}|10\ket + a_{11}|11\ket  
\nonumber 
\end{eqnarray} 
which involves one single-bit rotation $\alpha_1$ and 
two controlled$^1$-operations $U_{\alpha_{2,i}}$ ($i=0,1$) 
\begin{eqnarray} 
\alpha_1 &=& \arctan \sqrt{{{|a_{10}|^2  
+ |a_{11}|^2}}\over {{|a_{00}|^2 + |a_{01}|^2}}} \nonumber\\ 
U_{\alpha_{2,0}} &=& \left[\begin{array}{rr} 
  {a_{00}\over \sqrt{|a_{00}|^2+|a_{01}|^2}} & 
  {a_{01}\over \sqrt{|a_{00}|^2+|a_{01}|^2}} \\ 
  {a_{01}^*\over \sqrt{|a_{00}|^2+|a_{01}|^2}} & 
   -{a_{00}^*\over 
\sqrt{|a_{00}|^2+|a_{01}|^2}}\end{array}\right], \nonumber\\ 
U_{\alpha_{2,1}}& = &\left[\begin{array}{rr} 
  {a_{10}\over \sqrt{|a_{10}|^2+|a_{11}|^2}} & 
  {a_{11}\over \sqrt{|a_{10}|^2+|a_{11}|^2}} \\ 
  {a_{11}^*\over \sqrt{|a_{10}|^2+|a_{11}|^2}} & 
   -{a_{10}^*\over 
\sqrt{|a_{10}|^2+|a_{11}|^2}}\end{array}\right].   
\nonumber 
\end{eqnarray} 

The quantum  
circuit of a 3-qubit system transforms the state $|000\ket$ to 
an arbitrary superposed state with $N=2^3=8$ 
basis states. Starting from the $|000\ket$, a rotation with an angle 
$\arctan\sqrt{|a_{100}|^2+|a_{101}|^2+|a_{110}|^2+|a_{111}|^2 \over  
|a_{000}|^2+|a_{001}|^2+|a_{010}|^2+|a_{011}|^2}$ 
is operated on the 1st qubit, and this  
rotation transforms the state to 
$\sqrt{|a_{000}|^2+|a_{001}|^2+|a_{010}|^2+|a_{011}|^2}|000\ket$ +  
$\sqrt{|a_{100}|^2+|a_{101}|^2+|a_{110}|^2+|a_{111}|^2}|100\ket$. Then two 
controlled$^1$-rotations with angles 
$\arctan \sqrt{|a_{010}|^2+|a_{011}|^2 \over 
|a_{000}|^2+|a_{001}|^2}$ and $\arctan \sqrt{|a_{110}|^2+|a_{111}|^2 \over 
|a_{100}|^2+|a_{101}|^2}$ are applied to the 2nd qubit. 
The state vector becomes 
\begin{eqnarray} 
\sqrt{|a_{000}|^2+|a_{001}|^2}|000\ket&+
&\sqrt{|a_{010}|^2+|a_{011}|^2}|010\ket\nonumber\\ 
 + 
\sqrt{|a_{100}|^2+|a_{101}|^2}|100\ket&+
&\sqrt{|a_{110}|^2+|a_{111}|^2}|110\ket.\nonumber 
\end{eqnarray} 
Finally 4 controlled$^2$ unitary transformations 
\begin{eqnarray} 
U_{\alpha_{3,00}}&=&\left[\begin{array}{rr}{a_{000}\over \sqrt{|a_{000}|^2+|a_{001}|^2}}& 
{a_{001}\over \sqrt{|a_{000}|^2+|a_{001}|^2}}\\ 
{a^{*}_{001}\over \sqrt{|a_{000}|^2+|a_{001}|^2}}& 
-{a^{*}_{000}\over 
\sqrt{|a_{000}|^2+|a_{001}|^2}}\end{array}\right],\nonumber\\ 
U_{\alpha_{3,01}}&=&\left[\begin{array}{rr}{a_{010}\over \sqrt{|a_{010}|^2+|a_{011}|^2}}& 
{a_{011}\over \sqrt{|a_{010}|^2+|a_{011}|^2}}\\ 
{a^{*}_{011}\over \sqrt{|a_{010}|^2+|a_{011}|^2}}& 
-{a^{*}_{010}\over  
\sqrt{|a_{010}|^2+|a_{011}|^2}}\end{array}\right],\nonumber\\ 
U_{\alpha_{3,10}}&=&\left[\begin{array}{rr}{a_{100}\over \sqrt{|a_{100}|^2+|a_{101}|^2}}& 
{a_{101}\over \sqrt{|a_{100}|^2+|a_{101}|^2}}\\ 
{a^{*}_{101}\over \sqrt{|a_{100}|^2+|a_{101}|^2}}& 
-{a^{*}_{100}\over 
\sqrt{|a_{100}|^2+|a_{101}|^2}}\end{array}\right],\nonumber\\ 
U_{\alpha_{3,11}}&=&\left[\begin{array}{rr}{a_{110}\over \sqrt{|a_{110}|^2+|a_{111}|^2}}& 
{a_{111}\over \sqrt{|a_{110}|^2+|a_{111}|^2}}\\ 
{a^{*}_{111}\over \sqrt{|a_{110}|^2+|a_{111}|^2}}& 
-{a^{*}_{110}\over 
\sqrt{|a_{110}|^2+|a_{111}|^2}}\end{array}\right] \nonumber 
\end{eqnarray} 
are operated on the 3rd qubit to acquire the general 
superposed state $a_{000}|000\ket + 
a_{001}|001\ket$ +$a_{010}|010\ket + 
a_{011}|011\ket$ +$a_{100}|100\ket + a_{101}|101\ket$ 
+$a_{110}|110\ket + a_{111}|111\ket$. 
This quantum circuit is illustrated in Fig. 1. 
 
For brevity in notations, we use an ``angle" to label a 
controlled$^k$-operation. 
If the involved coefficients are all real, it reduces to an 
ordinary rotation angle. In the above notations for angles of  
the controlled$^k$-rotations, and in similar notations hereafter,  
the first subscript (for example, 3 in $\alpha_{3,01}$) 
refers to the target qubit order number 
and the following subscripts (01 in $\alpha_{3,01}$)  
indicate the quantum states of the controlling qubits. 
 
In the initialization, 
operations for the first $n-1$ qubits are controlled rotations 
where each rotation depends only on a single real parameter. 
The rotation angles    
take the following general expressions. 
In the 1st qubit,  
there is a 1-qubit rotation. The rotation angle is 
\begin{eqnarray} 
\alpha_1&=&\arctan\sqrt{\sum_{i_2 i_3 \cdots i_n}|a_{1i_2 i_3 \cdots i_n}|^2  
\over \sum_{i_2 i_3 \cdots i_n}|a_{0 i_2 i_3 \cdots i_n}|^2}. 
\label{first} 
\end{eqnarray} 
In the 2nd qubit, there are two controlled$^1$-rotations 
\begin{eqnarray} 
\alpha_{2, 0}&=&\arctan\sqrt{\sum_{i_3 i_4 \cdots i_n}  
|a_{01i_3 i_4 \cdots i_n}|^2 
\over \sum_{i_3 i_4 \cdots i_n}|a_{00i_3 i_4 \cdots i_n}|^2},\nonumber\\ 
\alpha_{2, 1}&=&\arctan\sqrt{\sum_{i_3i_4\cdots i_n} 
|a_{11i_3i_4\cdots i_n}|^2  
\over \sum_{i_3 i_4 \cdots i_n} |a_{10i_3 i_4 \cdots i_n}|^2}. 
\end{eqnarray} 
In general, in the $j$th qubit, there are $2^{j-1}$  
controlled$^{j-1}$-rotations, with each of them having  
$j-1$ controlling qubits labeled as $i_1i_2\cdots i_{j-1}$.  
The rotation angle in the $j$th qubit ($j\ne n$) is given by 
\begin{eqnarray} 
\alpha_{j, i_1i_2\cdots i_{j-1}}&=&\arctan\sqrt{  
\sum_{i_{j+1}\cdots i_n}|a_{i_1 i_2 
\cdots i_{j-1} 1  
i_{j+1}\cdots i_n}|^2 \over \sum_{i_{j+1}\cdots i_n}  
|a_{i_1 i_2\cdots i_{j-1} 0 
i_{j+1}\cdots i_n}|^2}. 
\label{e3} 
\end{eqnarray} 
The fraction in Eq. (\ref{e3}) can be ${0 \over 0}$ and the rotation 
angle in this case is  
undetermined. If this should happen,  
a simple analysis is 
sufficient for 
us to determine which gate operation should be adopted. 
Examples will be given later. 
 
For the last qubit with $j=n$, we have $2^{n-1}$ controlled$^{n-1}$ 
unitary transformations  
\begin{eqnarray} 
U_{\alpha_{n, i_1 i_2 \cdots i_{n-1}}} 
=\left[ 
\begin{array}{rr} {A_0 \over \sqrt{|A_0|^2+|A_1|^2}} &  
                  {A_1 \over \sqrt{|A_0|^2+|A_1|^2}} \\ 
                  {A_1^{\ast}\over \sqrt{|A_0|^2+|A_1|^2}} &  
                 -{A_0^{\ast}\over \sqrt{|A_0|^2+|A_1|^2}}   
\end{array}\right], 
\end{eqnarray} 
with  
\begin{eqnarray} 
A_0=a_{i_1 i_2 \cdots i_{n-1} 0},\nonumber\\ 
A_1=a_{i_1 i_2 \cdots i_{n-1} 1}. 
\label{last1} 
\end{eqnarray} 
If the numbers in Eq. (\ref{last1}) are real, 
the operation is just a usual rotation, and the angle is given by 
\begin{eqnarray} 
\alpha_{n, i_1 i_2\cdots i_{n-1}}=\arctan {a_{i_1 i_2\cdots i_{n-1} 1}  
\over a_{i_1 i_2  
\cdots i_{n-1} 0}}. 
\label{last} 
\end{eqnarray} 
 
In general, the $N$ amplitudes ${a_i}$ in Eq. (\ref{e1}) are complex,
and together with the normalization, the total number of real parameters 
for description of a superposed state is $2N-1$. 
In our scheme, 
operations on the first $n-1$ qubits are all ordinary rotations, 
and they provide  
$1+2+4+\cdots +2^{n-2}=N/2-1$ real parameters. 
The $N/2$ operations on the last qubit are  
generally unitary transformations, 
and each of them depends on 3 real parameters. 
Altogether, the total 
number of real parameters involved in the initialization is $2N-1$. 
This number can be  
reduced if the state has special properties. 
For instance, if all the amplitudes are 
real, the number is reduced to $N-1$. 
 
Our scheme requires only $N-1$ gate 
operations to initialize a quantum register.  
In terms of the standard 1- and 2-bit 
gate operations, the total number of operations is $O(N n^2)$ 
which is still polynomial in $N$. It is more 
than the number of steps ($O(N n)$) in the VM protocol \cite{ventura}. 
This is the price to be paid for saving $n+1$ qubits in the 
register. The present scheme uses $n$ qubits, 
whereas the VM protocol requires  
$2n+1$ qubits to perform a same task. Barenco {\it et al.
}\cite{bar} pointed out that introduction of one more qubit to workspace 
will reduce the number of 
controlled$^m$-gate operations from $O(m^2)$ to $O(m)$. 
According to this, the number would increase from $O(N n)$ to $O(N n^{n+2})$
if we want to save $n+1$ qubits. 
It is surprisingly seen that the actual number required in our protocol
is much less than the estimation. 
 
In many practical cases, the number of controlled gate  
operations can be reduced and the circuit is accordingly  
simplified. Fig. 2 shows an example for part of a circuit  
where the rotation angles  
are the same. 
In this case, one can combine the $|0\ket$-controlled Hadamard-Walsh 
transformation and the   
$|1\ket$-controlled Hadamard-Walsh transformation as one operation, which is 
equivalent to one  
Hadamard-Walsh transformation on the target qubit. 
Consequently, the four controlled operations are reduced to a 
single-qubit rotation.  
 
If desired superposition has a special form,  
the quantum circuit can likely be 
further  
simplified. Next,  
we discuss three well-known cases.  
Starting with $|0\cdots 0\ket$, we initialize quantum superpositions   
of  
1) the evenly 
distributed  
state; 
2) the GHZ state; 
and 3) the state vector $|\psi\ket=\sin\theta |\tau\ket +\cos\theta|c\ket$  
which is used in Grover's 
quantum search algorithm. 
 
1) The evenly distributed state $|\psi\ket=\sum_{i}|i\ket$ is widely used in 
quantum  
computation. The Hadamard-Walsh gate operation on each 
qubit  
generates this form of  
superposition from the state $|0\cdots 0\ket$. This is 
also true for our scheme.  
In this special case, all rotation angles in   
Eqs. (\ref{first} - \ref{last}) are  
$\pi\over 4$, and all  
gate operations are therefore the Hadamard-Walsh transformation. In each 
qubit, the  
controlling qubits exhaust all possible combinations, and hence the 
$2^{j-1}$  
controlled Hadamard-Walsh gate operations can be reduced to a single 
Hadamard-Walsh  
transformation in the $j$th qubit. 
 
2) The GHZ state \cite{GHZ} is the maximally entangled state with the form  
of superposition ${1\over 
\sqrt{2}} 
(|0\cdots 0\ket \pm |1\cdots 1\ket)$.  
An example that transforms $|0000\ket$ to   
${1\over \sqrt{2} }(|0000\ket + |1111\ket)$ is given in Fig. 3. 
It can be seen that the circuit is much simplified from the most general one 
in Fig. 1.  
According to Eqs. (\ref{first} - \ref{last}),  
the simplification is achieved through the following steps.  
The rotation  
in the 1st qubit is the Hadamard-Walsh transformation. For the two   
controlled  
rotations in the 2nd qubit, $\alpha_{2,0}=0$ means the identity operation which 
does nothing  
for the qubit, and $\alpha_{2,1}={\pi \over 2}$ corresponds to the controlled  
NOT operation. So effectively,   
there is only one controlled NOT gate in the 2nd qubit. There are originally  
four gate 
operations in  
the 3rd qubit. 
$\alpha_{3,11}={\pi \over 2}$ is the $|11\ket$-controlled NOT gate and 
$\alpha_{3,00}$ is the identity operation. 
$\alpha_{3,01}$ and $\alpha_{3,10}$  
are undetermined angles with ${0\over 
0}$.  
By analyzing this problem, it is easy to see that the angles  
should be 0, which   
corresponds to the identity operation. Therefore, there is only one gate 
operation in  
the 3rd qubit: $|11\ket$-controlled NOT. Similarly, there is only  
$|111\ket$-controlled  
NOT operation in the 4th qubit.  
If the circuit consists of more than four qubits,  
the same analysis applies till the last but one 
qubit. In  
the last qubit, the rotation is either $\pi\over 2$ for ${1\over \sqrt{2}} 
(|0\cdots 0\ket  
+|1\cdots 1\ket)$, or $-{\pi\over 2}$  
for ${1\over \sqrt{2}}(|0\cdots 0\ket - |1\cdots 1\ket)$. 
 
3) In Grover's quantum search algorithm \cite{r3} and its generalizations 
\cite{rlong}, the  
state vector is built in a two-dimensional space spanned   
by the marked state $|\tau\ket$ 
and the ``rest" state $|c\ket=\sum_{i\ne \tau}|i\ket$. At any search step, the 
state vector has the form  
$|\psi\ket=\sin\theta|\tau\ket +\cos\theta|c\ket$. We now give 
the quantum  
circuit for initializing such a superposed state.  
Let $|\tau\ket=|i_1 i_2  
\cdots i_n\ket$ be the marked state,  
and we now construct $|\psi\ket$ from $|0\cdots 0\ket$.  The 
amplitudes  
of the basis states in Eq. (\ref{e1}) are $a_{\tau}=\sin\theta$ and  
$a_i=\cos\theta/\sqrt{N-1}$ for $i\ne 
\tau$.  
According to Eq. (\ref{first}), the rotation angle in the 1st qubit  
is 
\begin{eqnarray} 
\alpha_1=\left\{ 
\begin{array}{ll}  
\arctan \Omega_1, & {\rm if}\;\; i_1=1\\  
\arctan {1\over \Omega_1}, & {\rm if}\;\; i_1=0  
\end{array}\right.  
\nonumber 
\end{eqnarray} 
with  
\begin{displaymath}  
\Omega_1=\sqrt{(N-2)\cos^2\theta + 2(N-1)\sin^2\theta 
\over N\cos^2\theta}.   
\end{displaymath}  
In the $k$-th qubit, the angle for $|i_1 i_2 ...i_{k-1}\ket$-controlled 
rotation is 
\begin{eqnarray} 
\alpha_{k,i_1 i_2 ...i_{k-1}}=\left\{ 
\begin{array}{ll} 
\arctan \Omega_k, & {\rm if}\;\; i_k=1\\  
\arctan {1\over \Omega_k}, & {\rm if}\;\; i_k=0  
\end{array}\right. 
\nonumber 
\end{eqnarray} 
with  
\begin{displaymath}  
\Omega_1=\sqrt{(N-2^k)\cos^2\theta + 2^k(N-1)\sin^2\theta 
\over N\cos^2\theta}.   
\end{displaymath}  
The other rotation angles in the $k$-th qubit are all equal to ${\pi \over 
4}$, corresponding to  
the Hadamard-Walsh transformation. Thus, the $2^{k-1}-1$ controlled gate 
operations are  
reduced to $k-1$ controlled Hadamard-Walsh transformations. In Fig. 4, we 
show an  
example with the marked state $|\tau\ket=|i_1i_2i_3i_4\ket$ in 
a 4-qubit-system. In this 
example,  
$\cos\theta$ and $\sin\theta$ are all real and positive.  
 
To summarize, we have presented a general scheme for 
initializing a quantum register to an arbitrary superposed state. The 
quantum circuits 
utilize only single-qubit rotations and controlled 
qubit  
rotations. General expressions for rotation angles  
have been derived explicitly,  
and possibility for simplifying the circuits has been discussed in terms of  
three well-known superposed states. 
 
Y.S. thanks the Department of Physics of Tsinghua University 
for warm hospitality, and for support from 
the senior visiting scholar program. 
This work is supported by the Major State Basic Research Developed Program 
Grant No. G2000077400, the China National Natural Science Foundation Grant 
No. 60073009, the Fok 
Ying Tung Education Foundation, and the 
Excellent Young University Teachers' Fund 
of Education Ministry of China.

\begin{figure} 
\caption{ 
Quantum circuit for initializing an arbitrary superposed state  
of Eq. (\protect\ref{e1}) for a 3-qubit-register. 
} 
\label{figure.1} 
\end{figure} 
 
\begin{figure} 
\caption{ 
Example of combining controlled-rotations to simplify the circuit.  
} 
\label{figure.2} 
\end{figure} 
 
\begin{figure} 
\caption{ 
Quantum circuit for implementing the GHZ state.   
The circled H represents the Hadamard-Walsh transformation,  
and $\oplus$ the controlled-NOT gate. 
} 
\label{figure.3} 
\end{figure} 
 
\begin{figure} 
\caption{ 
Quantum circuit for implementing the state vector   
$|\psi\ket=\sin\theta|\tau\ket+\cos\theta|c\ket$.  
In the figure, a letter with a bar 
indicates its NOT value, i.e. $\bar{1}=0$ and $\bar{0}=1$.  
} 
\label{figure.4} 
\end{figure} 
 
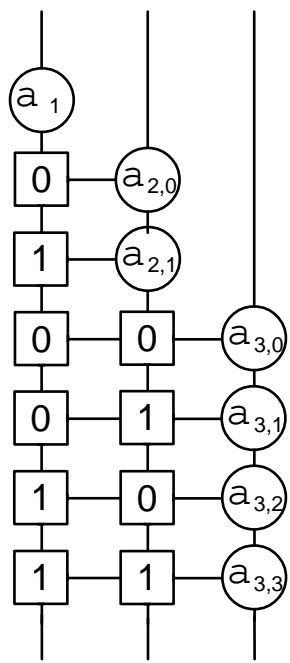
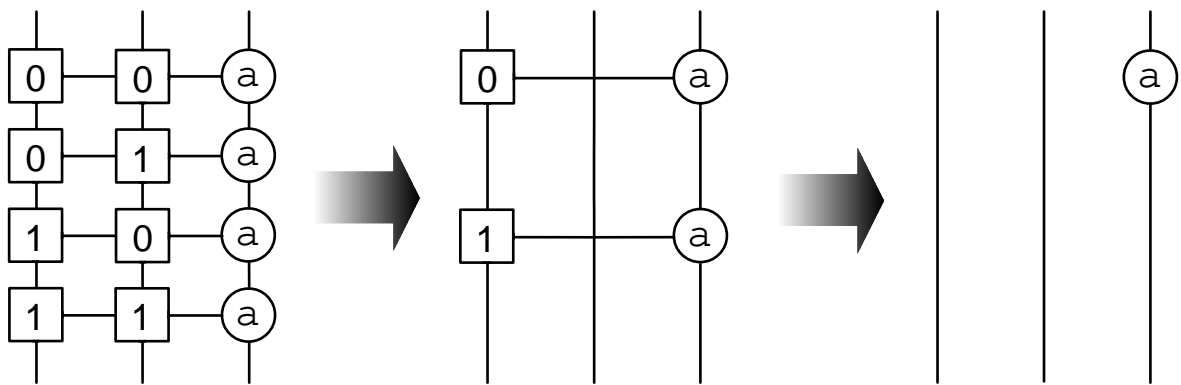
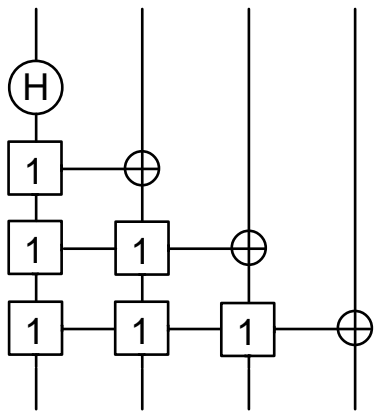
\end{document}